# High thermoelectric performance and low thermal conductivity of densified LaOBiSSe

Atsuhiro Nishida[1], Osuke Miura[1], Chul-Ho Lee[2], Yoshikazu Mizuguchi[1]*

*1. Department of Electrical and Electronic Engineering, Tokyo Metropolitan University, 1-1, Minami-osawa, Hachioji 192-0397, Japan*
*2. National Institute of Advanced Industrial Science and Technology, 1-1-1 Umezono, Tsukuba, Ibaraki 305-8568 Japan.*

Corresponding author: Yoshikazu Mizuguchi (mizugu@tmu.ac.jp)

In this study, we examined the thermoelectric properties of the layered bismuth-chalcogenide-based (BiCh$_2$-based) compound LaOBiSSe, which is expected to be a new candidate material for thermoelectric applications. Densified samples were obtained with a hot-press method. The results of the X-ray diffraction measurements showed that the samples obtained were weakly oriented. This affected the resistivity ($\rho$) and Seebeck coefficient ($S$) of the samples because they are dependent upon the orientation of the crystal structure. The values obtained for the power factor ($S^2/\rho$) when measured perpendicular to the pressing direction ($\mathbf{P}_\perp$) were higher than that when measured parallel to the pressing direction ($\mathbf{P}_{//}$). The thermal conductivity ($\kappa$) of the samples was also sensitive to the orientation. The values of $\kappa$ measured along $\mathbf{P}_{//}$ were lower than that measured along $\mathbf{P}_\perp$. The highest figure-of-merit (approximately 0.36) was obtained at around 650 K in both directions, i.e., $\mathbf{P}_\perp$ and $\mathbf{P}_{//}$.

The continuous improvement of the thermoelectric properties of materials is required for the development of various device-based applications.[1] One of the key strategies for improving thermoelectric properties is creating a layered structure. The low-dimensional electronic states that result from such two-dimensional crystal structures can be advantageous for producing high thermoelectric properties. It has been argued that quantum effects can also contribute to a high thermoelectric performance.[2,3] Several layered compounds, such as $Bi_2Te_3$, Co oxides, and $CsBi_4Te_6$, have exhibited high thermoelectric performances.[4-6] Layered materials are also more flexible when designing new materials. For example, the large number of potential combinations of elements results in a large variety of materials, as demonstrated by the number of Cu oxides and Fe-based layered superconductors.[7,8] It is also possible to create materials where each layer has a different function, e.g., alternating conducting and insulating layers. Thus, because of these aforementioned features, layered structures are suitable for designing new, high-performance, thermoelectric materials.

Recently, $BiS_2$-based materials have been intensively studied in the fields of condensed matter physics and chemistry because of the increasing interest in superconductors.[9,10] To date, superconductivity has been reported for ten different $BiS_2$-based compounds that were subjected to electron-carrier doping (see the review article by Mizuguchi[11] for further information on these materials). The number of variations in the crystal structure of the $BiS_2$-based family of compounds is advantageous for designing new thermoelectric materials. For example, there have been many reports on the thermoelectric properties of $BiS_2$-based and $BiS_2$-related compounds.[12-15] The authors of this paper have reported that the thermoelectric performance of the $BiS_2$-based compound $LaOBiS_2$ can be enhanced by partially substituting S with Se.[13] The dimensionless figure-of-merit ($ZT$) for the material $LaOBiS_{1.2}Se_{0.8}$ was found to be 0.17 at 743 K. However, our previous study on the $LaOBiS_{2-x}Se_x$ system was performed with pellets that were formed from powdered samples with a relative density of approximately 85%, and the $ZT$ values were roughly calculated from the room-temperature thermal conductivity of $LaOBiS_2$ (2 W/mK). To investigate the thermal transport and electrical properties of the $LaOBiS_{2-x}Se_x$ system precisely, the thermoelectric properties must be measured with densified samples. Therefore, in this study, we investigated the thermoelectric properties of densified samples of LaOBiSSe (i.e., $x = 1$ in the $LaOBiS_{2-x}Se_x$ system), which were fabricated with a hot-press method. We measured a very low thermal conductivity at approximately 650 K, and thus, obtained a $ZT$ value of 0.36. The anisotropy of the thermoelectric properties of the hot-pressed samples was also investigated.

Polycrystalline samples of LaOBiSSe were prepared according to the solid-state-reaction method described in Ref. 13. The samples were then densified with a hot-press (HP) instrument under a uniaxial pressure of approximately 50 MPa. The LaOBiSSe powders were sealed in a graphite capsule (diameter = 15 mm) for the HP annealing. The samples were then annealed at 700 °C for 1 h. Thus, a densified LaOBiSSe sample with a relative density greater than 99% was obtained. The as-prepared HP-LaOBiSSe sample was then cut into several pieces. The phase purity was determined with X-ray diffraction (XRD) using the θ-2θ method with Cu-Kα radiation. To determine the crystal orientation of the HP-LaOBiSSe sample, XRD measurements were performed with the scattering vector parallel ($P_{//}$) and perpendicular ($P_{\perp}$) to the pressing direction (see Fig. 1(d) for a schematic diagram of the $P_{//}$ and $P_{\perp}$ directions). The electrical resistivity ($\rho$) and Seebeck coefficient ($S$) were measured with a four-point probe method (ZEM-3, ADVANCE RIKO) from room temperature to approximately 650 K. The thermal diffusivity ($D$) and heat capacity ($C_p$) were measured with the laser-flash diffusivity method (LFA-457, Netzsch) from room temperature to approximately 650 K using a standard sample of pyroceram. The thermal conductivity ($\kappa$) was calculated from $D$, $C_p$, and the density of the sample (6.60 g/cm$^3$). The dimensions of the samples for the $\kappa$ measurements were 10×10 mm and a thickness of 1 mm. All of the thermoelectric measurements were performed in the directions of $P_{\perp}$ and $P_{//}$.

Fig. 1(a) shows the XRD pattern of the powdered HP-LaOBiSSe. Nearly all of the XRD peaks were indexed with the *P*4/*nmm* space group, except for the small concentration of a La$_2$O$_3$ impurity (7 mol%). To estimate the degree of crystal orientation in the HP-LaOBiSSe sample, the XRD measurements were performed with the scattering vector oriented parallel to the $\mathbf{P}_\perp$ and $\mathbf{P}_{//}$ directions (Figs. 1(b) and (c), respectively).  These XRD patterns exhibit differences in the relative peak intensities.  To qualitatively discuss the crystal orientation, we compared the (102) and (004) peaks.  The intensities of the (102) and (004) peaks are almost identical in the powdered sample (Fig. 1(a)).  When the scattering vector is parallel to $\mathbf{P}_\perp$, the intensity of (102) the peak is greater than that of the (004) peak, which indicates that the *a*-axis (*ab*-plane) has become slightly oriented along the $\mathbf{P}_\perp$ direction by the HP process (Fig. 1(b)).  In contrast, when the scattering vector is parallel to $\mathbf{P}_{//}$, the intensity of the (102) peak is lower than that of the (004) peak, which indicates that the *c*-axis has become oriented along the $\mathbf{P}_{//}$ direction by the HP process (Fig. 1(c)).  However, these differences are not significant, indicating that the as-prepared HP-LaOBiSSe sample is only weakly oriented.

We then measured the thermoelectric properties along the directions of $\mathbf{P}_\perp$ and $\mathbf{P}_{//}$.  Fig. 2(a) shows $\rho$ as a function of the temperature for the HP-LaOBiSSe sample when the current (*I*) was parallel to the $\mathbf{P}_\perp$ and $\mathbf{P}_{//}$ directions (i.e., $I // \mathbf{P}_\perp$ and $I // \mathbf{P}_{//}$, respectively).  For both directions, $\rho$ increases with increasing temperature.  However, the values of $\rho$ when $I // \mathbf{P}_\perp$ are lower than that when $I // \mathbf{P}_{//}$, which is caused by the enhanced electrical conductivity along the *a*-axis (in-plane conductivity).  Fig. 2(b) shows *S* as a function of the temperature for the HP-LaOBiSSe sample when the temperature difference (*ΔT*) was applied along the $\mathbf{P}_\perp$ and $\mathbf{P}_{//}$ directions (i.e., $\Delta T$ // $\mathbf{P}_\perp$ and $\Delta T$ // $\mathbf{P}_{//}$, respectively). Both of the directions exhibit negative values for $S$, indicating that the contributing charge-carriers are electrons. In addition, $S$ decreases (the absolute value of $S$ increases) with increasing temperature in both directions. The absolute value of $S$ for $\Delta T$ // $\mathbf{P}_\perp$ is slightly larger than that for $\Delta T$ // $\mathbf{P}_{//}$. These differences in the electrical properties are caused by the crystal orientation and anisotropic electrical conduction of $BiCh_2$-based compounds.[16]

Figs. 3(a) and (b) show the lattice thermal-conductivity ($\kappa_{lattice}$) and total thermal-conductivity ($\kappa_{total}$), respectively, as functions of the temperature for the HP-LaOBiSSe sample with $\Delta T$ // $\mathbf{P}_\perp$ and $\Delta T$ // $\mathbf{P}_{//}$. $\kappa_{lattice}$ is obtained by subtracting the charge-carrier thermal conductivity ($\kappa_{electron}$) from $\kappa_{total}$. $\kappa_{electron}$ can be estimated with the equation $\kappa_{electron} = LT/\rho$, where $L$ and $T$ are the Lorenz number and absolute temperature, respectively. Both $\kappa_{lattice}$ and $\kappa_{total}$ decrease with increasing temperature, irrespective of the direction of $\Delta T$. The $\kappa_{lattice}$ and $\kappa_{total}$ values when $\Delta T$ // $\mathbf{P}_{//}$ are significantly lower than that when $\Delta T$ // $\mathbf{P}_\perp$, indicating the anisotropy of $\kappa$ along the $\mathbf{P}_\perp$ and $\mathbf{P}_{//}$ directions. The anisotropy of $\kappa$ is caused by the crystal orientation in the as-prepared HP-LaOBiSSe sample.

The calculated power factor ($S^2/\rho$) as a function of the temperature, for both the $\mathbf{P}_\perp$ and $\mathbf{P}_{//}$ directions, is plotted in Fig. 4(a). The power factor in the direction of $\mathbf{P}_\perp$ is higher than that in the direction of $\mathbf{P}_{//}$. A high power factor of approximately 6 μW/K$^2$cm was obtained above 500

K in the direction of $\mathbf{P}_\perp$. Fig. 4(b) shows *ZT* as a function of the temperature for the HP-LaOBiSSe sample when measured along $\mathbf{P}_\perp$ and $\mathbf{P}_{//}$. *ZT* increases with increasing temperature in both directions. The *ZT* values in the direction of $\mathbf{P}_{//}$ are higher than that in the direction of $\mathbf{P}_\perp$ over the entire temperature range tested. The higher *ZT* values in the direction of $\mathbf{P}_{//}$ were attained because of the lower $\kappa$ values in this direction. At approximately 650 K, the highest *ZT* value (approximately 0.36) was obtained in both the $\mathbf{P}_\perp$ and $\mathbf{P}_{//}$ directions. However, *ZT* is still increasing at the highest temperature tested in this study, and thus, higher *ZT* values are expected at higher temperatures with the as-prepared HP-LaOBiSSe samples. Furthermore, we believe that the power factor and $\kappa$ can be further improved by substituting the conducting BiSSe layer and/or the insulating LaO layer. In addition, to understand the origin of the low $\kappa$, an analysis of the crystal structure at high temperatures is needed, and this research is currently being performed.

In conclusion, we investigated the thermoelectric properties of a densified HP-LaOBiSSe sample. A weak crystal oriented was revealed by XRD analysis. The power factor in the direction of $\mathbf{P}_\perp$ was higher than that in the direction of $\mathbf{P}_{//}$, which is caused by the enhanced in-plane electrical conductivity. The temperature dependence of $\kappa$ in the direction of both $\mathbf{P}_\perp$ and $\mathbf{P}_{//}$ was anisotropic. The $\kappa$ values measured in the direction of $\mathbf{P}_{//}$ were much lower than that measured in the direction of $\mathbf{P}_\perp$ direction. The low $\kappa$ values in the direction of $\mathbf{P}_{//}$ were caused by the low $\kappa$ values in the direction of the $c$-axis of the HP-LaOBiSSe crystal structure. The highest $ZT$ value of 0.36 was obtained in both measurement directions at approximately 650 K.

Acknowledgements

The authors thank H. Nishiate, M. Kunii, M. Aihara, and A. Yamamoto Y. Goto for their technical supports.  This work was partly supported by Grant-in-Aid for Young Scientist (A) (25707031) and research fund from TEET.

References

[1] G. J. Snyder, and E.S.Toberer, Nat. Mater. **7**, 105 (2008).

[2] L. D. Hicks, and M. S. Dresselhaus, Phys. Rev. B **47**, 12727 (1993).

[3] J. P. Heremans, V. Jovovic, E. S. Toberer, A. Saramat, K. Kurosaki, A. Charoenphakdee, S. Yamanaka, and G. J. Snyder, Science **321**, 554 (2008).

[4] H. J. Goldsmid, proc. Phys. Soc. London **71**, 633 (1958).

[5] I. Terasaki, Y. Sasago, and K. Uchinokura, Phys. Rev. B **56**, 12685(R) (1997).

[6] D. Y. Chung, T. Hogan, P. Brazis, M. R. Lane, C. Kannewurf, M. Bastea, C. Uher, and M. G. Kanatzidis, Science **287**, 1024 (2000).

[7] J. B. Bednorz and K. Müller, Z. Physik B Condensed Matter **64**, 189-193 (1986).

[8] Y. Kamihara et al., J. Am. Chem. Soc. **130**, 3296–3297 (2008).

[9] Y. Mizuguchi, H. Fujihisa, Y. Gotoh, K. Suzuki, H. Usui, K. Kuroki, S. Demura, Y. Takano, H. Izawa, and O. Miura, Phys. Rev. B. **86**, 220510 (2012).

[10] Y. Mizuguchi, S. Demura, K. Deguchi, Y. Takano, H. Fujihisa, Y. Gotoh, H. Izawa, and O. Miura, J. Phys. Soc. Jpn. **81**, 114725 (2012).

[11] Y. Mizuguchi, J. Phys. Chem. Solid **84**, 34–48 (2015).

[12] A. Omachi, J. Kajitani, T. Hiroi, O. Miura, and Y. Mizuguchi, J. Appl. Phys. **115**, 083909 (2014).

[13] Y. Mizuguchi, A. Omachi, Y. Goto, Y. Kamihara, M. Matoba, T. Hiroi, J. Kajitani, and O. Miura, J. Appl. Phys. **116**, 163915 (2014).

[14] Y. L. Sun, A. Ablimit, H. F. Zhai, J. K. Bao, Z. T. Tang, X. B. Wang, N. L. Wang, C. M. Feng, and G. H. Cao, Inorg. Chem. **53**, 11125 (2014).

[15] Y. Goto, J. Kajitani, Y. Mizuguchi, Y. Kamihara, and M. Matoba, J. Phys. Soc. Jpn. **84**, 085003 (2015).

[16] M. Nagao, A. Miura, S. Watauchi, Y. Takano, I. Tanaka, Jpn. J. Appl. Phys. **54**, 083101 (2015).

Fig. 1.

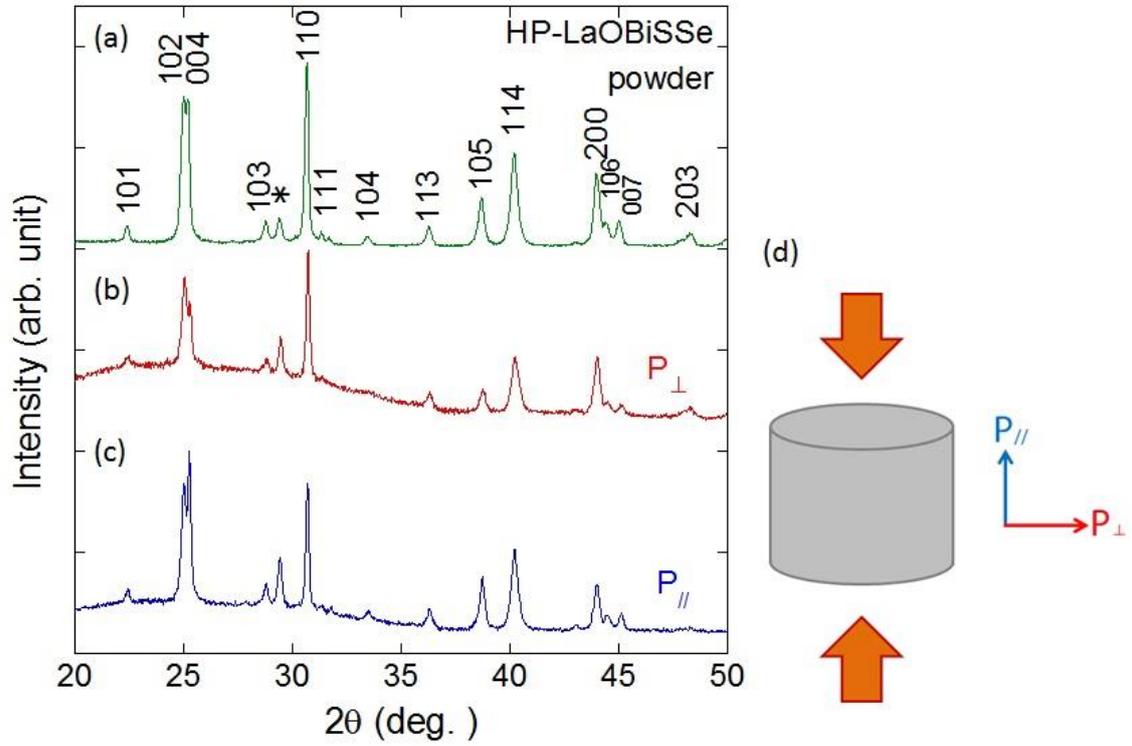

Fig. 1. (a) XRD pattern of powdered HP-LaOBiSSe. (b) XRD pattern of HP-LaOBiSSe measured perpendicular to the press direction ($P_\perp$). (c) XRD pattern of HP-LaOBiSSe measured parallel to the press direction ($P_{//}$). (d) Schematic image of the definition of the $P_\perp$ and $P_{//}$ directions with the pressing direction in HP.

Fig. 2.

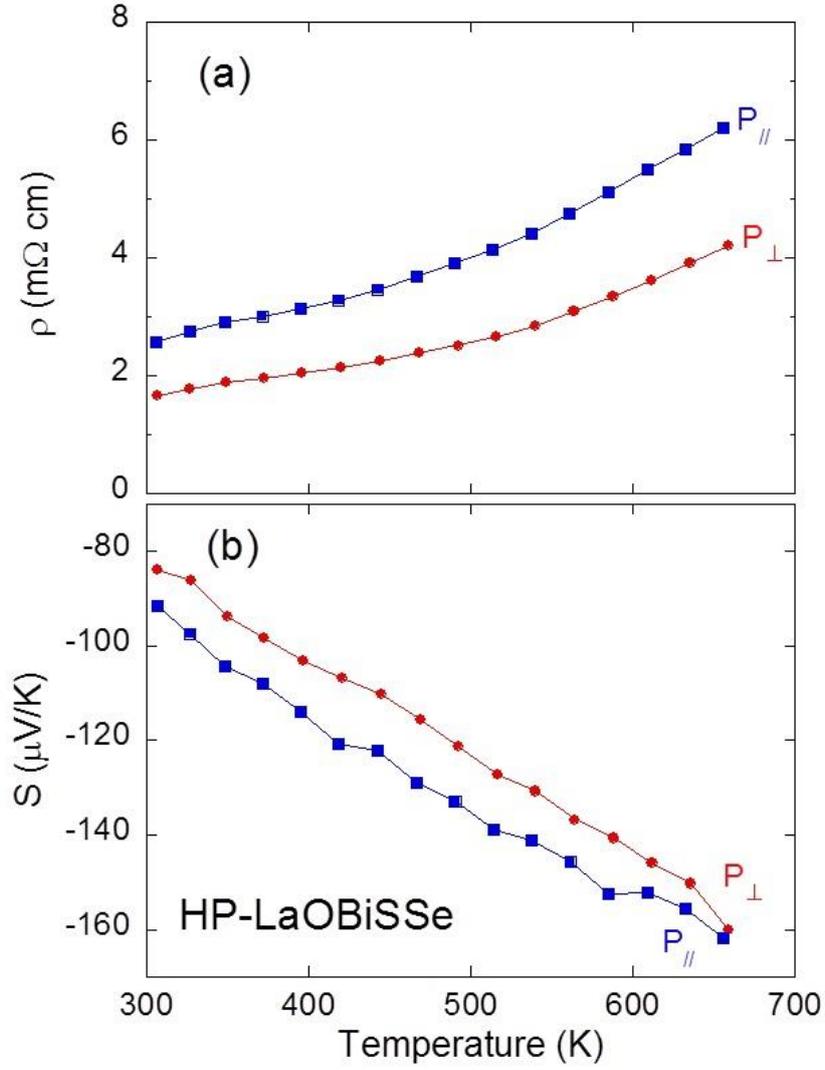

Fig. 2. (a) Temperature dependences of $\rho$ for HP-LaOBiSSe with current direction of $I$ // $\mathbf{P}_\perp$ and $I$ // $\mathbf{P}_{//}$. (b) Temperature dependences of $S$ for HP-LaOBiSSe with $\Delta T$ // $\mathbf{P}_\perp$ and $\Delta T$ // $\mathbf{P}_{//}$.

Fig. 3.

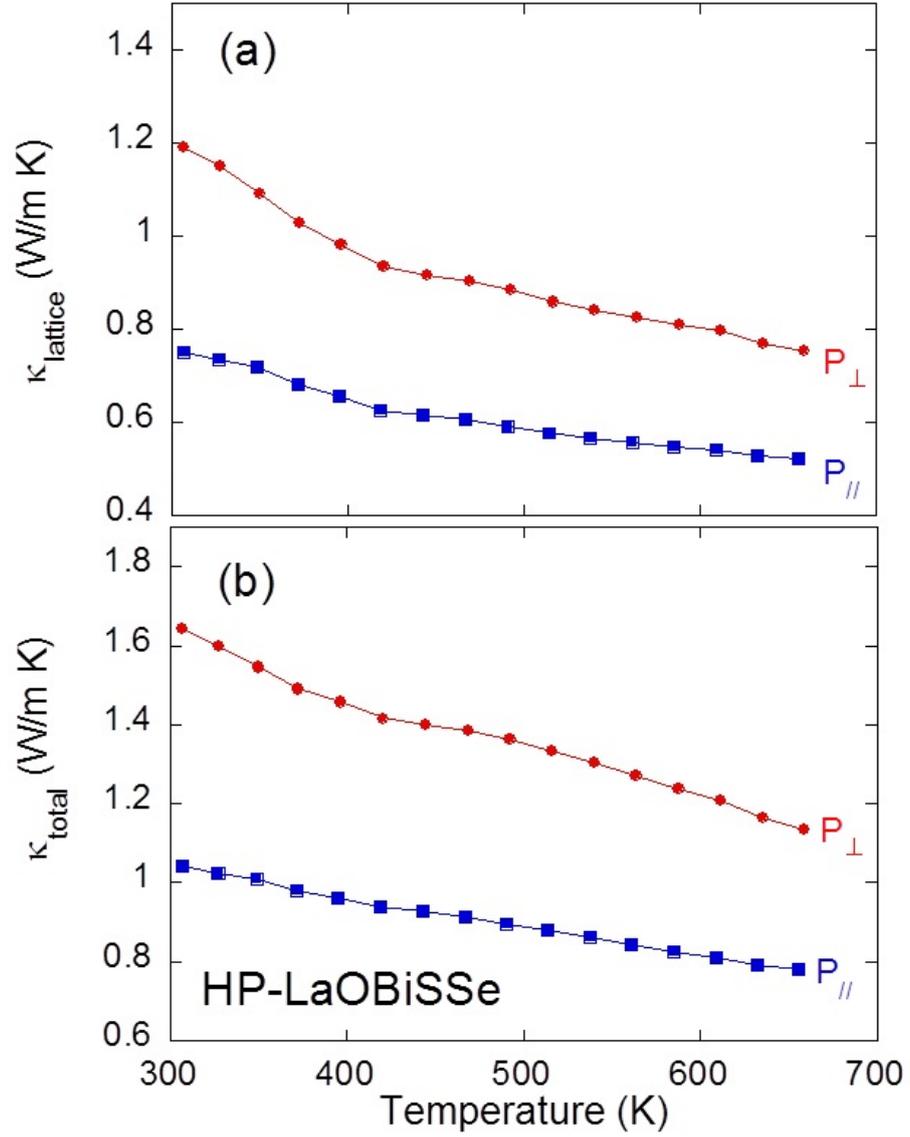

Fig. 3. (a) Temperature dependences of $\kappa_{lattice}$ for HP-LaOBiSSe with $\Delta T$ // $\mathbf{P}_\perp$ and $\Delta T$ // $\mathbf{P}_{//}$.

(b) Temperature dependences of $\kappa_{total}$ for HP-LaOBiSSe with $\Delta T$ // $\mathbf{P}_\perp$ and $\Delta T$ // $\mathbf{P}_{//}$.

Fig. 4.

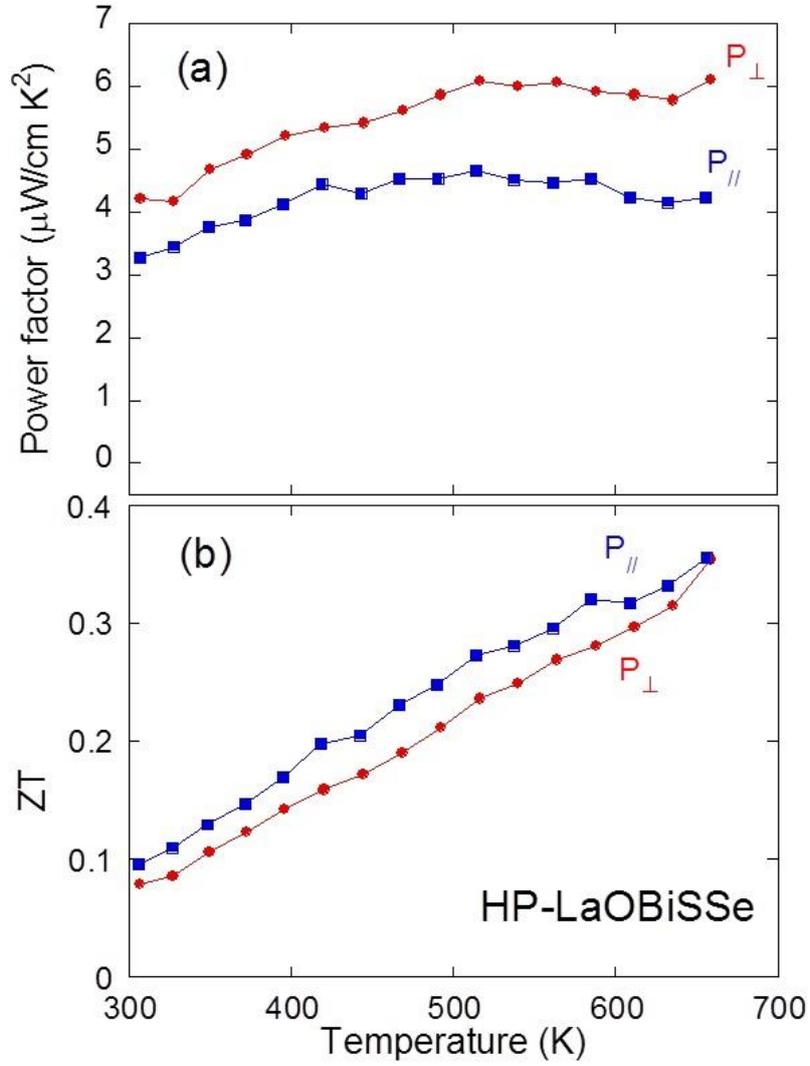

Fig. 4. (a) Temperature dependences of power factor for HP-LaOBiSSe with two different directions in measurements ($P_\perp$ and $P_{//}$). (b) Temperature dependences of *ZT* for HP-LaOBiSSe with two different directions in measurements ($P_\perp$ and $P_{//}$).

14